\documentclass[aps,amsmath,amssymb,twocolumn]{revtex4-1}

\usepackage{graphicx}
\usepackage{dcolumn}
\usepackage{bm}
\usepackage{color}
\usepackage[normalem]{ulem}

\begin{document}

\title{Magnetic field effects in the near-field radiative heat transfer between planar structures}

\author{E. Moncada-Villa$^{1}$}
\author{J.~C. Cuevas$^{2}$}

\affiliation{$^{1}$Escuela de F\'{\i}sica, Universidad Pedag\'ogica y Tecnol\'ogica de Colombia,
Avenida Central del Norte 39-115, Tunja, Colombia}

\affiliation{$^2$Departamento de F\'{\i}sica Te\'orica de la Materia Condensada
and Condensed Matter Physics Center (IFIMAC), Universidad Aut\'onoma de Madrid,
E-28049 Madrid, Spain}

\date{\today}

\begin{abstract}
One of the main challenges in the field of thermal radiation is to actively control the near-field
radiative heat transfer (NFRHT) between closely spaced bodies. In this context, the use of an external 
magnetic field has emerged as a very attractive possibility and a plethora of physical phenomena have 
been put forward in the last few years. Here, we predict some additional magnetic-field-induced phenomena 
that can take place in the context of NFRHT between planar layered structures containing magneto-optical 
(MO) materials (mainly doped semiconductors like InSb). In particular, we predict the possibility of 
increasing the NFRHT upon applying an external magnetic field in an asymmetric structure consisting of 
two infinite plates made of InSb and Au. We also study the impact of a magnetic field in the NFRHT between 
structures containing MO thin films and show that the effect is more drastic than in their bulk counterparts.
Finally, we systematically investigate the anisotropic thermal magnetoresistance, i.e., the dependence of
the radiative heat conductance on the orientation of an external magnetic field, in the case of 
two infinite plates made of InSb and show that one can strongly modulate the NFRHT by simply changing
the orientation of the magnetic field. All the phenomena predicted in this work can be experimentally 
tested with existent technology and provide a new insight into the topic of active control of NFRHT.
\end{abstract}

\maketitle

\section{Introduction}

The field of thermal radiation has been revolutionized in recent years by the confirmation of the prediction 
that the radiative heat transfer between closely spaced objects can largely overcome the Planckian or 
black-body limit, for recent reviews see Refs.~\cite{Song2015a,Cuevas2018}. Traditionally, our understanding of 
thermal radiation has been based on Planck's law and the concept of a black body. This law establishes,
in particular, an upper limit for the amount of heat that two bodies can exchange via electromagnetic 
radiation (Stefan-Boltzmann's law). However, Planck's law, which is based on ray optics, has obvious 
limitations and, in particular, fails to describe the radiative heat transfer between two objects when 
they are separated by distances below the thermal wavelength $\lambda_{\rm Th}$ ($\sim$10 $\mu$m at room 
temperature). This was first recognized in the early 1970s by Polder and van Hove \cite{Polder1971}, who 
made use of the theoretical framework of fluctuational electrodynamics derived in the 1950s by Rytov 
\cite{Rytov1953,Rytov1989} to predict that the radiative heat transfer between two infinite parallel plates 
could great overcome the black-body limit by bringing them sufficiently close (that is, below $\lambda_{\rm Th}$). 
They showed that in this near-field regime the radiative heat transfer can be entirely dominated by evanescent 
waves, which give an extra contribution that is not taken into account in Stefan-Boltzmann's law. The 
NFRHT enhancement was already hinted in several experiments in the late 1960's \cite{Hargreaves1969,Domoto1970}, 
but due to experimental challenges it took more than 30 years to confirm it unambiguously. In the last
decade or so, different experiments have demonstrated the possibility to overcome the black-body limit
in the near-field regime by exploring a large variety of materials and body shapes  
\cite{Kittel2005,Narayanaswamy2008,Hu2008,Rousseau2009,Shen2009,Shen2012,Ottens2011,Kralik2012,
Zwol2012a,Zwol2012b,Guha2012,Worbes2013,Shi2013,St-Gelais2014,Song2015b,Kim2015,Lim2015,St-Gelais2016,
Song2016,Bernardi2016,Cui2017,Kloppstech2017,Ghashami2018,Fiorino2018,DeSutter2019}.
Moreover, these experiments have confirmed the validity of the theory of fluctuational electrodynamics 
and they have also triggered off the hope that NFRHT may have an impact in different thermal technologies, 
see Refs.~\cite{Song2015a,Cuevas2018}.

Now that, to a large extent, the basic principles governing NFRHT have been established, one of the central challenges 
is to learn how to actively control thermal emission in the near-field regime, something that would be essential
for its use in a variety of applications. In this context, the possibility of using an external magnetic field, 
mimicking somehow what is being done in spintronics, has emerged as one of the most interesting ideas.
The use of an external magnetic field to control the NFRHT between magneto-optical (MO) materials
was first put forward in our work of Ref.~\cite{Moncada-Villa2015}. There, we showed that the NFRHT
between two parallel plates made of doped semiconductors can be strongly affected by the application of 
a static magnetic field and relative changes of up to 700\% can be induced with fields of a few Teslas 
\cite{Moncada-Villa2015}. Ever since, a plethora of thermomagnetic effects have been predicted. Thus,
for instance, it has been suggested that the lack of reciprocity in MO systems can lead to novel 
phenomena such as a near-field thermal Hall effect \cite{Ben-Abdallah2016a} or the existence of a 
persistent heat current \cite{Zhu2016}. It has also been shown that MO materials under a static magnetic 
field can exhibit the near-field thermal analogues of the key effects in the field of spintronics such as
a giant thermal magnetoresistance \cite{Latella2017} or an anisotropic thermal magnetoresistance 
\cite{Abraham-Ekeroth2018}. Some of these phenomena and other ones have been reviewed in Ref.~\cite{Ott2019}.

Although the fundamental importance of the thermomagnetic effects mentioned above is unquestionable,
many of them deal with submicron particles and many-body systems, which are very challenging to 
explore experimentally. Fortunately, there are still many interesting open questions related to the
magnetic field dependence of the NFRHT between planar structures containing MO materials. Thus, for 
instance, in our work of Ref.~\cite{Moncada-Villa2015} on the NFRHT between infinite plates made of 
doped semiconductors, an ideal class of MO materials, we always found that the magnetic field reduces
the radiative heat conductance, irrespective of its magnitude and direction. Thus, a natural 
question arises on whether it is possible to enhance the NFRHT upon applying an external magnetic
field. Another interesting question is related to the magnetic field dependence of the NFRHT between
structures containing thin films of MO materials. As we discovered in the case of doped semiconductors, 
the application of a magnetic field in these materials leads to the appearance of hyperbolic modes 
\cite{Moncada-Villa2015,Moncada-Villa2019}, very much like in hyperbolic metamaterials where the diagonal
components of the permittivity tensor have different signs \cite{Poddubny2013}. These hyperbolic modes where 
shown to play a very important role in the field-dependent NFRHT between doped semiconductors 
\cite{Moncada-Villa2015}, and since they are propagating waves, they are expected to be very 
sensitive to the presence of substrates in structures containing thin films. So, in this respect, one wonders
what is the magnetic field dependence of the NFRHT in thin-film structures of MO materials. On the other hand,
it has been recently shown in the context of submicron MO particles that the NFRHT is very sensivite
to the orientation of the magnetic field with respect to the transport direction \cite{Abraham-Ekeroth2018}. 
The variation of the radiative heat conductance with the orientation of the external field gives rise
to thermal analog of the anisotropic magnetoresistance (AMR) in spintronics and it has been referred to
as anisotropic thermal magnetoresistance (ATMR) \cite{Abraham-Ekeroth2018}. In that work, the field modulation 
of the NFRHT in submicron InSb particles was found to be orders of magnitude larger than in the corresponding
electrical conductance in AMR devices. Moreover, the ATMR phenomenon was suggested to have potential 
applications in the field of ultrafast thermal management as well as magnetic and thermal remote sensing. 
Given the fundamental interest in this phenomenon, it would be highly desirable to analyze it in the 
context of macroscopic planar structures, where it would be much easier to investigate experimentally.
The goal of this work is to tackle all these open questions from the theoretical point of view and to 
provide very concrete predictions that could potentially be tested with existent experimental techniques.          

The rest of the paper is organized as follows. Section \ref{sec-theory} introduces the systems under 
study and presents the general formalism that we use for the description of NFRHT in the presence of
a magnetic field. In Sec.~\ref{sec-InSb-Au} we study the NFRHT between two parallel plates made of InSb 
and, and we show that it is possible to enhance it by applying an external magnetic field. Section 
\ref{sec-thin-film} is devoted to the analysis of the magnetic field dependence of the NFRHT between 
structures containing thin films of MO materials. Section \ref{sec-ATMR} presents a detailed discussion
of the anisotropic thermal magnetoresistance in a system comprising two infinite parallel plates made
of InSb. Finally, Sec.~\ref{sec-conclusions} summarizes the main conclusions of our work.
 
\section{Systems under study and theoretical approach} \label{sec-theory}

\begin{figure}[t]
\includegraphics[width=\columnwidth,clip]{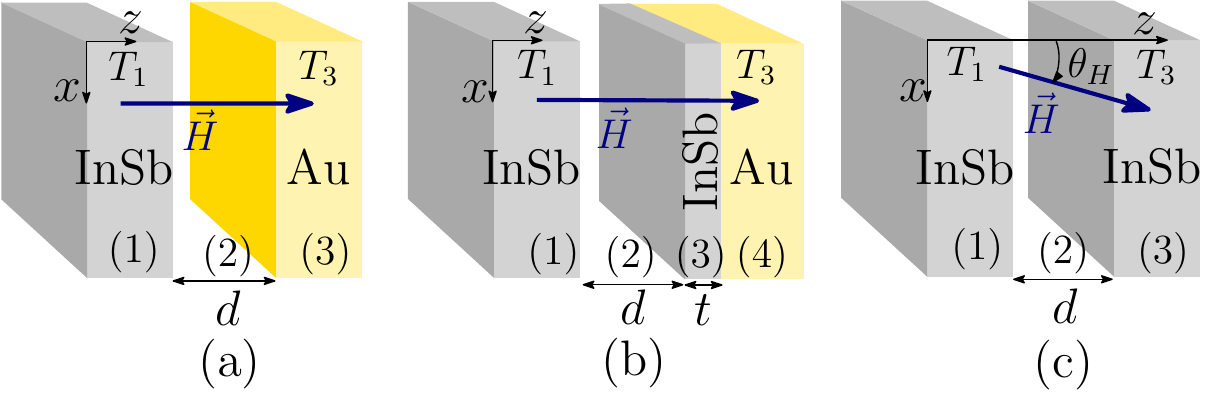}
\caption{Schematic representation of the systems under study. (a) Two semi-infinite plates made of 
\textit{n}-doped InSb and Au at temperatures $T_1$ and $T_3$, respectively, separated by vacuum gap 
of size $d$ and subjected to an external magnetic field that is perpendicular to the plates. (b) A 
semi-infinite plate of \textit{n}-doped InSb at temperature $T_1$ separated by a vacuum gap (of size 
$d$) from a InSb thin film of thickness $t$, deposited on an Au semi-infinite substrate. The thin 
film and the substrate have the same temperature $T_3$. (c) Two semi-infinite plates made of 
\textit{n}-doped InSb at temperatures $T_1$ and $T_3$, respectively, separated by vacuum gap of size 
$d$ and subjected to an external magnetic field in the $xz$ that forms an angle $\theta_H$ with the 
$z$-direction. In all three cases the vacuum gap is referred to as medium 2.}
\label{fig-system}
\end{figure}

Our main goal is to compute the radiative heat transfer between different layered systems in 
the presence of an external dc magnetic field, see Fig.~\ref{fig-system}, within the framework of 
fluctuational electrodynamics \cite{Rytov1953,Rytov1989}. All the systems considered here consist
of two layered systems separated by a vacuum gap (medium 2) of size $d$ and subjected to an 
external magnetic field, $\vec{H}$, that can point in any direction. In general, the layer 
materials can have MO activity, i.e., under the presence of a magnetic field these
materials are optically anisotropic materials and their permittivity can be described by a tensor 
of the form \cite{Zvezdin1997}
\begin{equation}
\label{perm-tensor}
\hat \epsilon = \left( \begin{array}{ccc}
\epsilon_{xx} & \epsilon_{xy} & \epsilon_{xz} \\
\epsilon_{yx} & \epsilon_{yy} & \epsilon_{yz} \\
\epsilon_{zx} & \epsilon_{zy} & \epsilon_{zz} \end{array} \right) .
\end{equation}
Here, as we indicate in Fig.~\ref{fig-system}, $x$ and $y$ lie in the planes of the layer interfaces 
and $z$ corresponds to the surface normal. The components of this permittivity tensor depend, in general,
on the applied magnetic field, as we shall specify below, and on the frequency (local approximation). 
Let us remind that the off-diagonal elements in Eq.~(\ref{perm-tensor}) are responsible for the polarization
conversion and all the typical MO effects (Faraday effect, Kerr effects, etc.) \cite{Zvezdin1997}.
Thus, our problem is to compute the radiative heat transfer between two anisotropic planar layered systems. 
This generic problem has been addressed before in the literature \cite{Biehs2011,Moncada-Villa2015} and we 
just recall here the central result. The net power per unit of area exchanged between two anisotropic layered 
systems is given by the following Landauer-like expression \cite{Biehs2011,Moncada-Villa2015}
\begin{equation}
\label{eq-net-Q}
Q = \int^{\infty}_{0} \frac{d \omega}{2\pi} \left[ \Theta_1(\omega) - \Theta_3(\omega) 
\right] \int \frac{d{\bf k}}{(2\pi)^2} \tau(\omega,{\bf k},d) ,
\end{equation}
where $\Theta_i(\omega) = \hbar \omega/ [\exp(\hbar \omega / k_{\rm B}T_i) -1]$, $T_i$ is the absolute 
temperature of the layer $i$, $\omega$ is the radiation frequency, ${\bf k} = (k_x,k_y)$ is the wave 
vector parallel to the surface planes, and $\tau(\omega,{\bf k},d)$ is the total transmission 
probability of the electromagnetic propagating waves ($|{\bf k}|=k < \omega/c$), as well as evanescent 
ones ($k> \omega/c$). This transmission coefficient is expressed as \cite{Biehs2011,Moncada-Villa2015}
\begin{eqnarray}
\label{eq-trans-man}
\tau(\omega,{\bf k},d) = \hspace{7cm} & & \\ \left\{ \begin{array}{ll}
\mbox{Tr} \left\{ [\hat 1 - \hat {\cal R}_{21} \hat {\cal R}^{\dagger}_{21} ] \hat {\cal D}^{\dagger}
[\hat 1 - \hat {\cal R}^{\dagger}_{23} \hat {\cal R}_{23} ] \hat {\cal D} \right\}, & k < \omega/c \\
\mbox{Tr} \left\{ [\hat {\cal R}_{21} - \hat {\cal R}^{\dagger}_{21} ] \hat {\cal D}^{\dagger}
[\hat {\cal R}^{\dagger}_{23} - \hat {\cal R}_{23} ] \hat {\cal D} \right\} e^{-2|q_2|d}, & k > \omega/c
\end{array} \right. , & & \nonumber
\end{eqnarray}
where $q_2 = \sqrt{\omega^2/c^2 - k^2}$ is the $z$-component of the wave vector in the vacuum gap, 
$c$ is the velocity of light in vacuum and $\hat {\cal D}= [ \hat 1 - \hat {\cal R}_{21} 
\hat {\cal R}_{23} e^{2iq_2d} ]^{-1}$ describes the usual Fabry-P\'erot-like denominator resulting 
from the multiple scattering between the two interfaces. The $2 \times 2$ matrices $\hat {\cal R}_{ij}$ 
are the reflections matrices characterizing the two interfaces at both sides of the gap. These matrices 
have the following generic structure
\begin{equation}
\label{refl-mat}
\hat {\cal R}_{ij} = \left( \begin{array}{cc} r^{s,s}_{ij} & r^{s,p}_{ij} \\ 
r^{p,s}_{ij} & r^{p,p}_{ij} \end{array} \right) ,
\end{equation}
where $r^{\alpha, \beta}_{ij}$ with $\alpha,\beta =s,p$ (or TE, TM) is the reflection amplitude for the 
scattering of an incoming $\alpha$-polarized plane wave into an outgoing $\beta$-polarized wave.
In practice, we compute numerically the different reflection matrices appearing in 
Eq.~(\ref{eq-trans-man}) by using the scattering-matrix approach for anisotropic multilayer systems
that is described in Refs.~\cite{Caballero2012,Moncada-Villa2015}. 

Throughout this work we focus on the analysis of the radiative linear heat conductance per unit
of area, $h$, which is known as the heat transfer coefficient. This coefficient is given by
\begin{equation}
h(T,d) = \lim_{\Delta T \rightarrow 0^+} \frac{Q(T_1=T+\Delta T,T_3=T,d)}{\Delta T} . 
\end{equation}
Additionally, we define the spectral heat flux, $h_{\omega}$, as the heat transfer coefficient per unit of 
frequency. Moreover, in all the calculations reported in this work we shall assume room temperature ($T=300$ K).

\begin{figure}[t]
\includegraphics[width=0.8\columnwidth,clip]{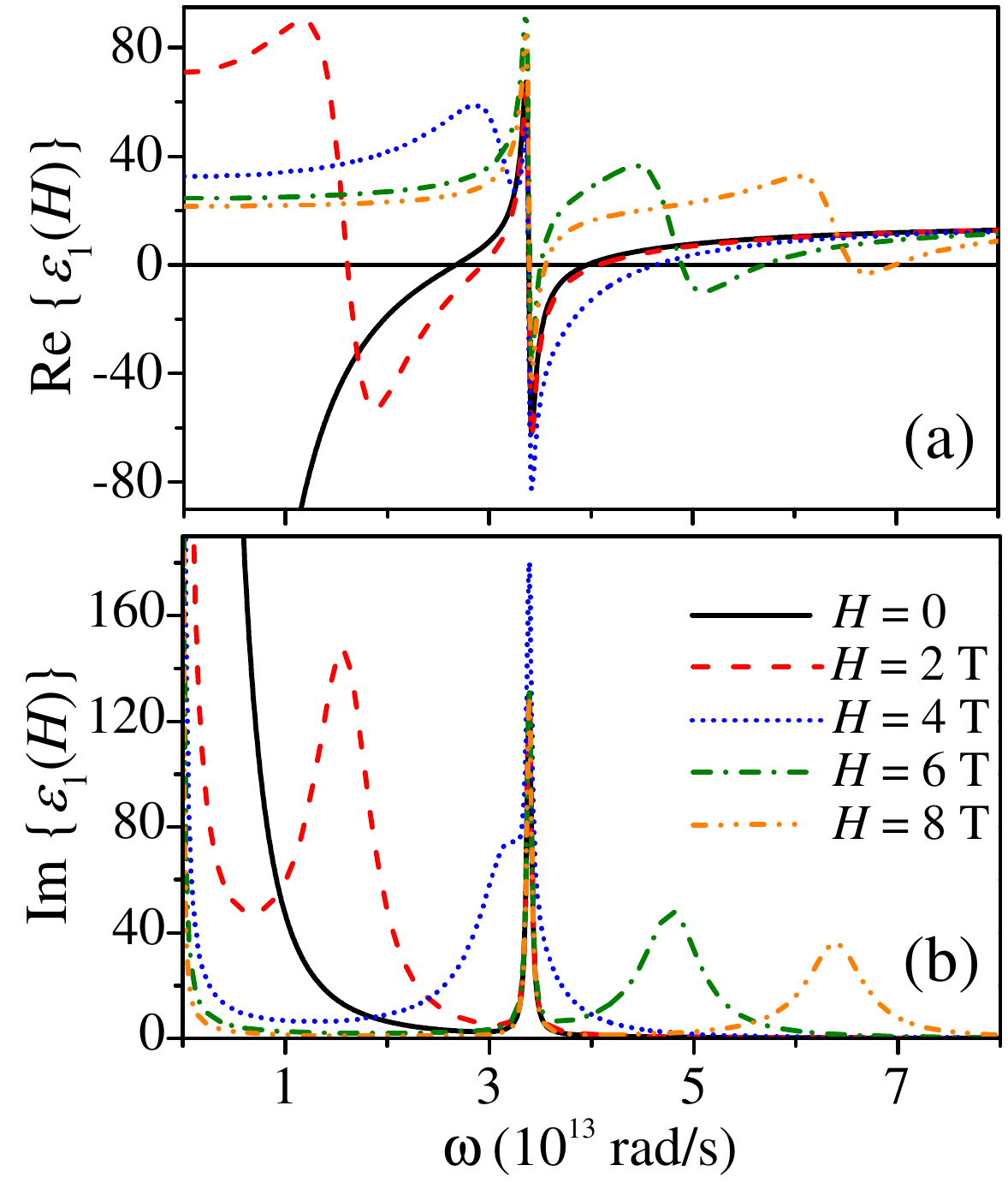}
\caption{(a) Real and (b) imaginary parts of the dielectric function $\epsilon_1(H)$, see 
Eq.~(\ref{eq-epsilons}), as a function of the frequency for different values of the magnitude 
of the external magnetic field. We remind that $\epsilon_1(H=0)=\epsilon_3$.}
\label{fig-dielectric}
\end{figure}

In the following sections we shall apply the general formalism above to the systems depicted in 
Fig.~\ref{fig-system}. All these systems contain at least one layer made of $n$-doped InSb under 
the action of a magnetic field in the $xz$ plane and forming an angle $\theta_H$ with the $z$-axis. 
As a consequence, the InSb can exhibit an optical anisotropy that can be described by the following 
permittivity tensor \cite{Palik1976}
\begin{widetext} 
\begin{equation}
\label{perm-tensor-theta}
\hat \epsilon = 
\left( \begin{array}{ccc} 
\epsilon_1 \cos^2\theta_H + \epsilon_3 \sin^2\theta_H &  -i\epsilon_2\cos\theta_H & 
\frac{1}{2}(\epsilon_1-\epsilon_3) \sin2\theta_H \\
i\epsilon_2\cos\theta_H   &   \epsilon_1   &   i\epsilon_2 \sin\theta_H \\
\frac{1}{2}(\epsilon_1-\epsilon_3) \sin2\theta_H & -i\epsilon_2 \sin\theta_H & 
\epsilon_1 \sin^2\theta_H + \epsilon_3 \cos^2\theta_H
\end{array} \right) ,
\end{equation}
\end{widetext}
with
\begin{eqnarray}
\epsilon_1(H) & = & \epsilon_{\infty} \left( 1 + \frac{\omega^2_L - \omega^2_T}{\omega^2_T - 
\omega^2 - i \Gamma \omega} + \frac{\omega^2_p (\omega + i \gamma)}{\omega [\omega^2_c -
(\omega + i \gamma)^2]} \right) , \nonumber \\
\label{eq-epsilons}
\epsilon_2(H) & = & \frac{\epsilon_{\infty} \omega^2_p \omega_c}{\omega [(\omega + i \gamma)^2 -
\omega^2_c]} , \\
\epsilon_3 & = & \epsilon_{\infty} \left( 1 + \frac{\omega^2_L - \omega^2_T}{\omega^2_T -
\omega^2 - i \Gamma \omega} - \frac{\omega^2_p}{\omega (\omega + i \gamma)} \right) . \nonumber
\end{eqnarray}
Here, $\epsilon_{\infty}$ is the high-frequency dielectric constant, $\omega_L$ ($\omega_T$ ) is 
the longitudinal (transverse) optical phonon frequency, $\omega^2_p =ne^2/(m^{\ast} \epsilon_0 
\epsilon_{\infty})$ is the plasma frequency of free carriers of density $n$ and effective mass 
$m^{\ast}$, $\Gamma$ ($\gamma$) is the phonon (free-carrier) damping constant, and $\omega_c = 
eH/m^{\ast}$ is the cyclotron frequency, which depends on the intensity of the applied magnetic 
field. For the sake of concreteness, we shall assume throughout this work that $\epsilon_{\infty} 
= 15.7$, $\omega_L = 3.62 \times 10^{13}$ rad/s, $\omega_T = 3.39\times 10^{13}$ rad/s, 
$\Gamma = 5.65 \times 10^{11}$ rad/s, $\gamma = 3.39 \times 10^{12}$ rad/s, $n = 1.07 \times 10^{17}$ 
cm$^{-3}$, $m^{\ast}/m = 0.022$, and $\omega_p = 3.14 \times 10^{13}$ rad/s \cite{Palik1976}. As a 
reference for our discussions below, we show in Fig.~\ref{fig-dielectric} the real and the
imaginary part of the function $\epsilon_1(H)$ for different values of the magnetic field in the
frequency range of relevance for the radiative heat transfer problems studied in this work. At zero
field, InSb is an optically isotropic material and $\epsilon_1$ is its dielectric function. In this
case, one can distinguish three frequency regions. First, for frequencies around $\omega_T$, the 
frequency of the transversal phonon, $\epsilon_1(H=0)$ exhibits the typical Lorentz-like behavior 
associated to an optical phonon with a sign change in the real part and a peak in the imaginary part.
In the region below $\omega_T$, $\epsilon_1(H=0)$ follows the typical Drude-like behavior of a metal, 
and above $\omega_T$ it behaves as a non-lossy dielectric. At a finite magnetic field, 
the real part of $\epsilon_1(H)$ exhibits additional sign changes, in particular, around the 
cyclotron frequency which is equal to $\omega_c = 8.02 \times 10^{12}$ rad/s for a field of 1 T. 
Notice also the appearance of the corresponding peak in the imaginary part at $\omega_c$. Let us
also recall that $\epsilon_3 = \epsilon_1(H=0)$. Thus, one can see that there are regions at
finite external field in which the real parts of $\epsilon_1$ and $\epsilon_3$, which mainly
determine the diagonal components of the permittivity tensor, have opposite signs. This fact
implies the appearance of hyperbolic modes, as it has been discussed before 
\cite{Moncada-Villa2015,Moncada-Villa2019}.

On the other hand, we shall also consider layers made of Au. For this metal, we shall use the
following Drude-like relative permittivity \cite{Chapuis2008}
\begin{equation}
\epsilon_{\rm{Au}}=\epsilon_{\infty,\rm{Au}}- \frac{\omega^2_{p,\rm{Au}}}
{\omega (\omega + i \gamma_{\rm{Au}})},
\end{equation}                           
where $\epsilon_{\infty,\rm{Au}}=1$, $\omega_{p,\rm{Au}}=1.71 \times 10^{16}$ rad/s and 
$\gamma_{\rm{Au}}=1.22 \times 10^{14}$ rad/s. We remark that we shall ignore the effect of
the magnetic field in the gold permittivity, which is justified by the huge plasma frequency
as compared to the cyclotron frequency for realistic values of the magnetic field.

\section{Magnetic field induced enhancement of the NFRHT} \label{sec-InSb-Au}

In previous studies of the radiative heat transfer between two MO objects,
it has always been found that the application of an external magnetic field reduces the 
linear heat conductance, irrespective of the orientation of the field. For instance, we
showed in Ref.~\cite{Moncada-Villa2015} that this was the case in parallel plates made of
identical materials (doped InSb and doped Si), for both the near-field and the far-field regime.
Something similar was found in Ref.~\cite{Abraham-Ekeroth2018} in the case of two identical 
spherical particles made of InSb of arbitrary size. However, there is no fundamental reason why 
the magnetic field should always reduce the radiative heat transfer between two bodies and 
the goal of this section is to show that indeed an external field can enhance, in particular,
the NFRHT between two objects. For this purpose, we analyze in this section the asymmetric system 
of Fig.~\ref{fig-system}(a) which consists of two infinite parallel plates made of doped InSb
(a MO material) and Au. For the sake of concreteness, we shall assume that the external magnetic 
field is applied perpendicularly to the plates. In this particular case, the magnetic-field-dependent 
permittivity tensor of InSb adopts the form  
\begin{equation}
\label{perm-tensor-Hz}
\hat \epsilon = 
\left( \begin{array}{ccc} 
\epsilon_1(H)   &  -i\epsilon_2(H)  &  0 \\
i\epsilon_2(H)  &   \epsilon_1(H)   & 0  \\
0               &         0         & \epsilon_3 \\
\end{array} \right) ,
\end{equation}
where the different elements are given by Eq.~(\ref{eq-epsilons}) and the diagonal components
are shown in Fig.~\ref{fig-dielectric} for different values of the magnetic field.

\begin{figure}[t]
\includegraphics[width=0.8\columnwidth,clip]{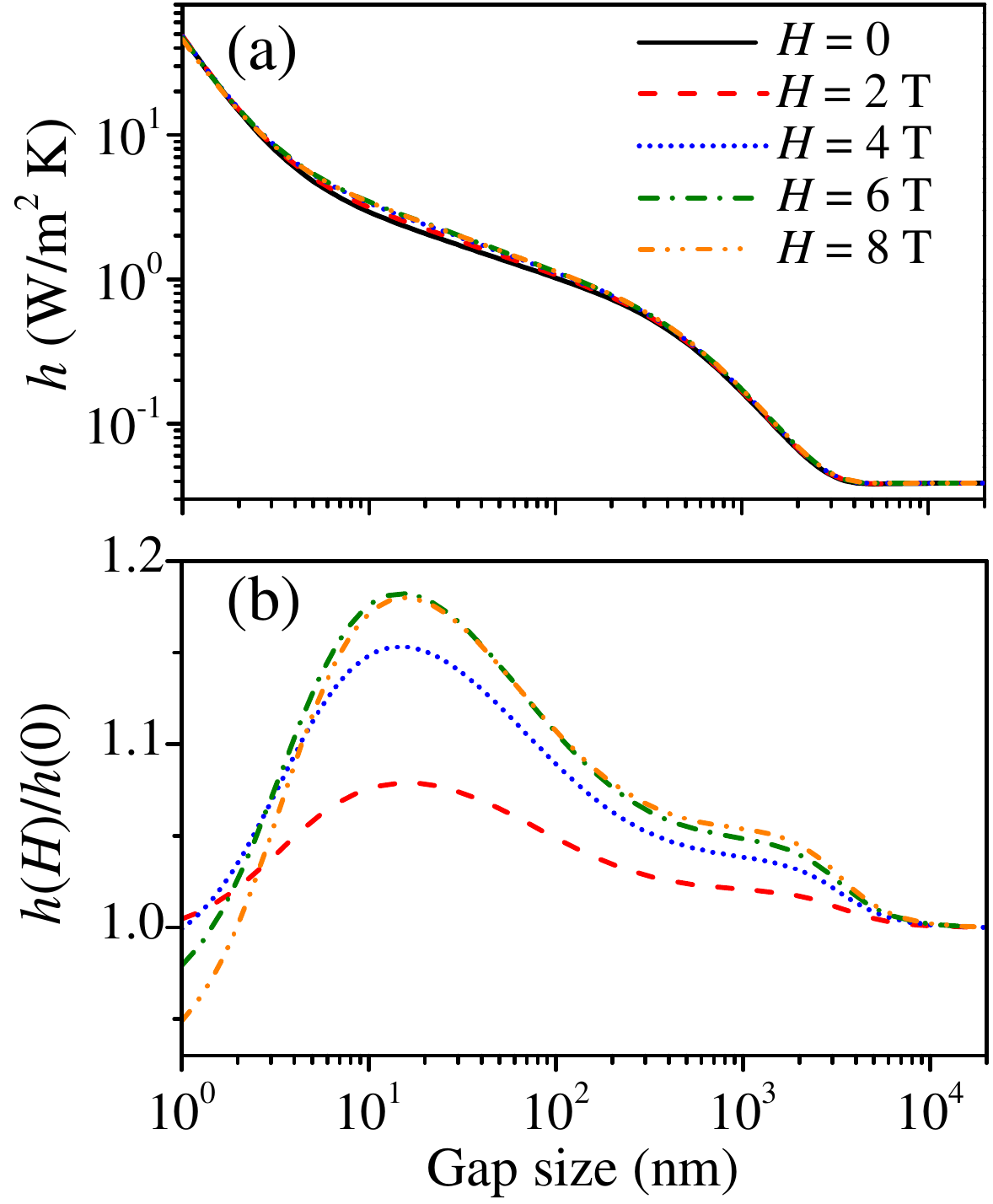}
\caption{(a) Heat transfer coefficient for the system of Fig.~\ref{fig-system}(a) 
as a function of the gap size for different values of the magnetic field perpendicular to the plate 
surfaces. (b) The corresponding ratio between the heat transfer coefficient at a given value of the 
field magnitude and the zero-field coefficient as a function of the gap size.}
\label{fig-InSb-Au-HTC}
\end{figure}

In Fig.~\ref{fig-InSb-Au-HTC}(a) we summarize the results for the heat transfer coefficient as 
a function of the separation between the InSb and the Au plates for different values of
the external magnetic field, which is applied perpendicularly to the plates. The first thing
to notice is that, as expected, the heat transfer coefficient is clearly lower than for the symmetric 
cases InSb-InSb \cite{Moncada-Villa2015} and Au-Au \cite{Chapuis2008}, irrespective of the gap 
size. This is simply due to the mismatch between the dielectric functions of both materials.
Actually, the heat conductance only overcomes the black-body limit (6.124 W m$^{-2}$ K$^{-1}$)
for very small gap below 10 nm. More importantly, the magnetic field enhances the conductance in
practically all the near-field regime (apart from very small gaps below 10 nm). This can be
clearly seen in Fig.~\ref{fig-InSb-Au-HTC}(b) where we show the ratio between the heat transfer
coefficient with and without magnetic field. Notice that the conductance is enhanced by more than 
20\% for fields of 6-8 T for gaps between 10 and 100 nm. Notice also that in the far-field 
regime (beyond 10 $\mu$m), the magnetic field barely changes the radiative heat conductance.

\begin{figure}[t]
\includegraphics[width=\columnwidth,clip]{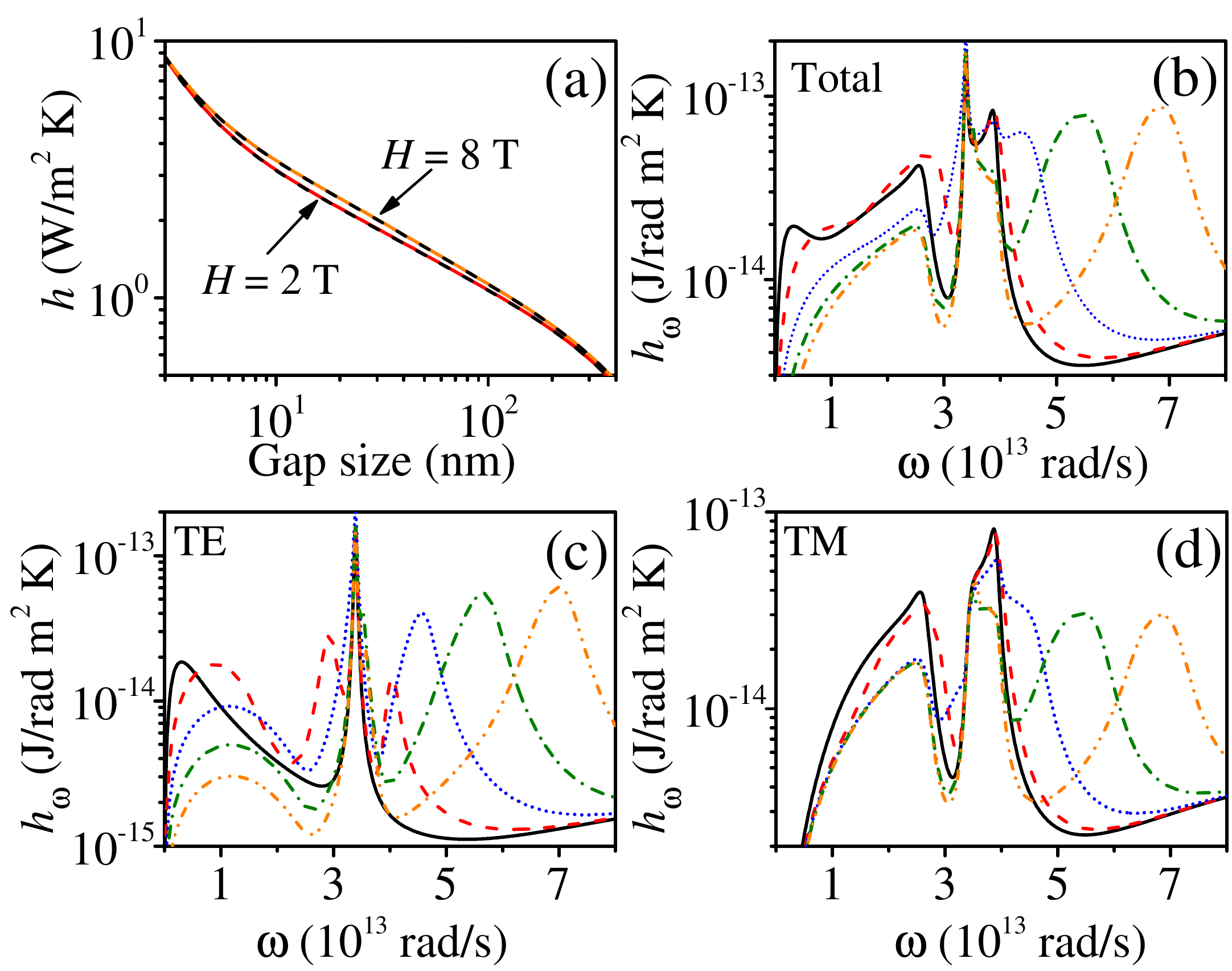}
\caption{ (a) Heat transfer coefficient as a function of the gap size for the system of 
Fig.~\ref{fig-system}(a) for a perpendicular magnetic field with magnitude of 2 and 8 T. The solid
lines correspond to the exact results and the dashed lines to the results obtained by considering 
no polarization conversion, that is, with $r^{s,p}_{ij} = r^{p,s}_{ij} = 0$ in Eq.~\eqref{refl-mat}. 
(b) The corresponding spectral heat flux computed with the exact formalism as a 
function of the frequency for a gap $d =10$ nm and different values of the magnitude of the 
external (perpendicular) field. The magnetic field values corresponding to the different lines are 
those indicated in the legend of Fig.~\ref{fig-InSb-Au-HTC}(a). In panels (c) and (d), we present, 
within the approximation $r^{s,p}_{ij} = r^{p,s}_{ij} = 0$, the spectral heat flux as a function of 
frequency for TE ($s$-polarized) and TM ($p$-polarized) waves, respectively, for a gap $d = 10$ nm 
and different values of the magnitude of the external (perpendicular) field.}
\label{fig-InSb-Au-spectrum}
\end{figure}

What is the physical origin for this magnetic-field-induced enhancement of the NFRHT? Let us recall
that in the symmetric case InSb-InSb it was found that the field reduces the NFRHT because the 
$p$-polarized (TM) surface modes that dominate the heat transfer in the near-field regime at zero 
field are progressively replaced by hyperbolic modes, which turn out to be less efficient transferring 
the heat across the gap \cite{Moncada-Villa2015}. In this case, the situation is clearly different.
First, the gap dependence that one can see in Fig.~\ref{fig-InSb-Au-HTC}(a) in the range $10$-$1000$
nm, where the conductance enhancement is most notable, suggests that the NFRHT is not dominated by surface 
modes, but rather by frustrated total internal reflection modes, like in the Au-Au case \cite{Chapuis2008}.
These modes are evanescent in the vacuum gap, but they are propagating inside the materials. Moreover,
in this case, and because of the off-diagonal elements in the permittivity tensor of Eq.~(\ref{perm-tensor-Hz}),
there is polarization conversion, whose role is not clear a priori and impedes to properly define the 
separate contributions of TE and TM modes. For these reasons, and in order to shed light on the 
physical interpretation, we have tested the validity of an approximation where we ignore the role of 
polarization conversion by setting the off-diagonal reflection coefficients to zero, i.e., we set 
$r^{s,p}_{ij} = r^{p,s}_{ij} = 0$ in Eq.~\eqref{refl-mat}. We stress that this is different from 
assuming that the off-diagonal elements of the InSb permittivity tensor vanish. Actually, those
elements play a non-negligible role in the diagonal reflection coefficients ($r^{s,s}_{ij}$ and
$r^{p,p}_{ij}$). As we show in Fig.~\ref{fig-InSb-Au-spectrum}(a) for two different values of the 
magnetic field, the approximation neglecting polarization conversion accurately reproduces the exact 
results. This fact allows us to conclude that the field-induced enhancement is not due to the polarization 
conversion, but rather to a modification of the probability of the evanescent waves. In particular, in 
the range where the near-field conductance is enhanced by the field, such an enhancement turns out to 
be mainly due to the enhanced probability of the TE modes, as we now proceed to show.

\begin{figure*}[t]
\includegraphics[width=\textwidth,clip]{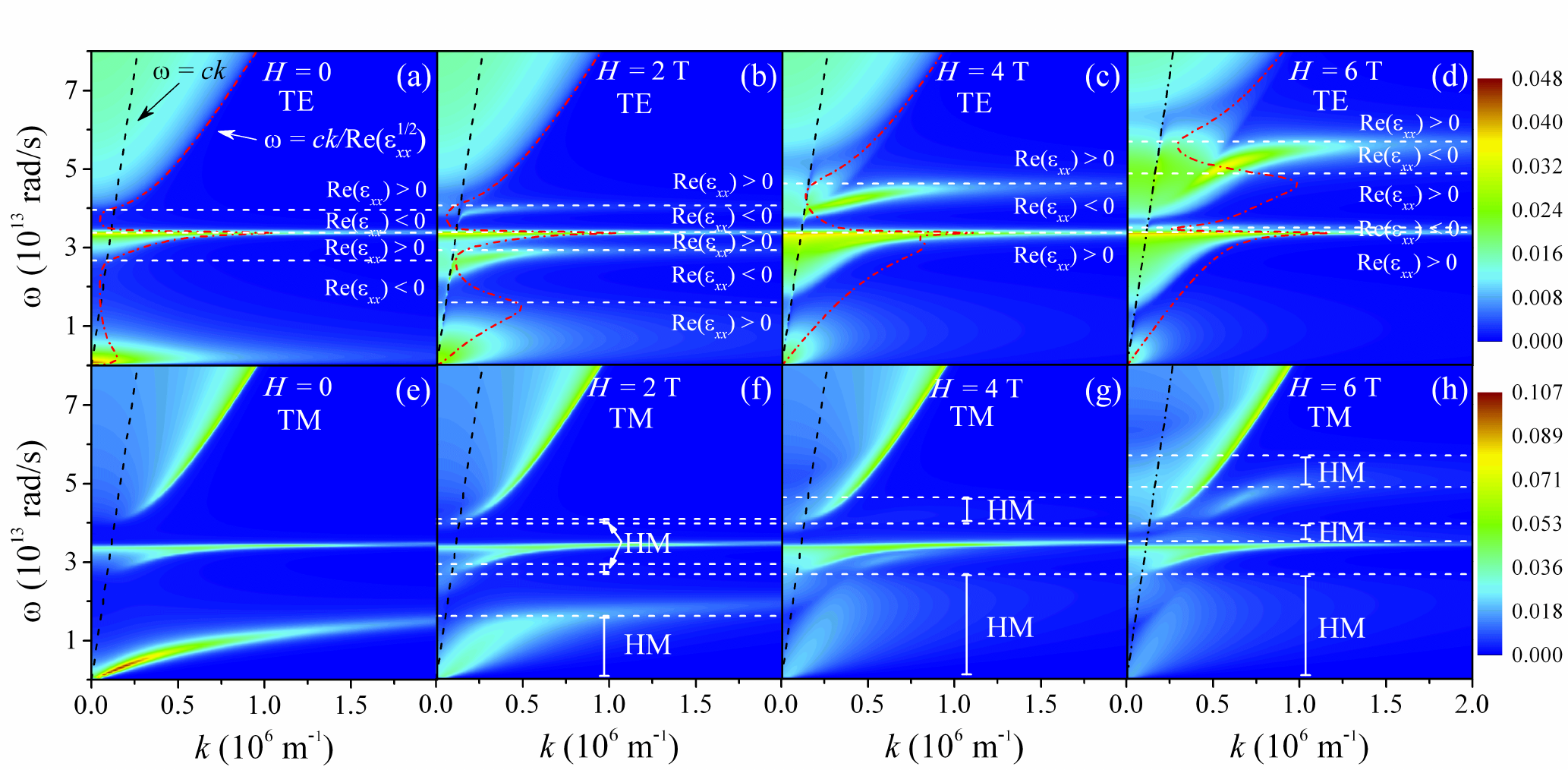}
\caption{Transmission coefficient for $s$-polarized (TE) [panels (a-d)] and $p$-polarized (TM) 
[panels (e-h)] waves, as a function of the magnitude of the parallel wave vector and frequency
for the system of Fig.~\ref{fig-system}(a) with a gap size $d=10$ nm. These transmissions were 
computed with the approximation neglecting polarization conversion and the different panels
correspond to different values the external (perpendicular) field. The black dashed lines correspond 
to the light line in vacuum, $\omega=ck$, while the red dashed-dotted lines in panels (a-d) correspond 
to the light line inside InSb, $\omega =ck \mbox{Re}(\epsilon^{1/2}_{xx})$. The horizontal dashed lines 
delimit the different regions where hyperbolic modes (HM) exist for the TM polarization and in the upper 
panels we specify the sign of $\mbox{Re} \{\epsilon_1(H) \}$ in these different regions.}
\label{fig-trans}
\end{figure*}

To illustrate that the TE modes are responsible for the field-induced enhancement, we first analyze the 
spectral heat flux. The total contribution to this quantity calculated with the exact formalism 
for a gap $d=10$ nm is shown in Fig.~\ref{fig-InSb-Au-spectrum}(b) as a function of the frequency
and for different values of the magnetic field. The corresponding contributions from TE and TM
modes calculated within the approximation $r^{s,p}_{ij} = r^{p,s}_{ij} = 0$ are shown in panels (c) 
and (d) of that figure, respectively. Although both contributions change with the applied field, the 
enhancement is clearly due to the TE modes. For this polarization (TE), the zero-field spectral heat 
flux is dominated by a peak at the frequency of the optical phonon $\omega_T = 3.39\times 10^{13}$ rad/s. 
As the field increases, there appear additional peaks in the spectral function at positions that
are intimately related to the frequencies at which the real part of $\epsilon_1(H)$ changes sign. 
In particular, notice the presence of a peak that follows closely, but not exactly, cyclotron frequency,
which proportional to the field and is equal to $8.02 \times 10^{12}$ rad/s for a field of 1 T. 
For the TM polarization, the main contributions at zero field to the spectral heat flux come from 
the Reststrahlen band (between $\omega_T$ and $\omega_L$) due to surface phonon polaritons in InSb 
and at low frequencies due to surface plasmon polaritons in InSb. At finite fields, there are additional 
contributions coming mainly from frequencies around the cyclotron frequency.

To further clarify the nature of the relevant electromagnetic modes we analyze their corresponding 
transmission coefficients (ignoring polarization conversion). These coefficients are shown in Fig.~\ref{fig-trans}
as a function of the frequency and the magnitude of the parallel wave vector for a gap of 10 nm. 
Focusing on the transmission of the TE modes (upper panels), we see that at zero field the main
contributions come from modes that are evanescent in the vacuum gap, but they are propagating 
inside the InSb plate (notice that they lie on the left of the light line inside InSb). In other 
words, these are frustrated total internal reflection modes. At finite magnetic field, the real
part of $\epsilon_{xx}=\epsilon_1$ exhibits additional sign changes and, in particular, there are
several regions where the real parts of $\epsilon_{xx}$ and $\epsilon_{zz}=\epsilon_3$ have 
opposite signs, see Fig.~\ref{fig-dielectric}. This latter fact means that the InSb can exhibit 
hyperbolic modes in those frequency regions for TM polarization, as it was amply discussed in 
Ref.~\cite{Moncada-Villa2015}. These regions with hyperbolic modes are denoted by HM in the lower 
panels of Fig.~\ref{fig-trans}. However, the dispersion relation of the ordinary waves for TE 
polarization is solely determined by $\epsilon_{xx}=\epsilon_1(H)$ and the new peaks in the spectral 
heat flux for this polarization, see Fig.~\ref{fig-InSb-Au-spectrum}(c), appear at frequencies at which 
the real part of $\epsilon_{xx}$ vanishes, while its imaginary part remains relatively appreciable. As one
can see in the upper panels of Fig.~\ref{fig-trans}, the main change induced by the external field
around those frequencies is the enhanced contribution of evanescent waves with larger values of the 
parallel wave vector, which naturally increases the total contribution to the heat transfer. 
The enhancement of the contribution of these modes is due to reduction of the impedance mismatch
between InSb and the vacuum gap, which in turn is due to the vanishing value of $\epsilon_{xx}$.
This reduction of the impedance mismatch is also signaled by the fact that around those frequencies
the light lines inside InSb and in vacuum tend to approach each other, see upper panels of 
Fig.~\ref{fig-trans}. With respect to the TM modes, see Fig.~\ref{fig-trans}(e-h), they are 
less sensitive to the external magnetic field and, in spite of the appearance of hyperbolic
modes, their contribution to the heat transfer continues to be dominated by frequencies inside 
the Reststrahlen band.

So, in short, we have seen in this section that an external magnetic field can indeed enhance 
the NFRHT in an asymmetric situation comprising an InSb and an Au plate. We have traced back the 
conductance enhancement in the near-field regime to an increase in the contribution of the
evanescent TE waves that overcomes the deleterious effect of the magnetic field on the evanescent
TM waves due to the appearance of hyperbolic modes.

\section{Magnetic field dependence of the radiative heat transfer in thin films} \label{sec-thin-film}

Recently, it has been experimentally demonstrated that thin films made of polar dielectrics may 
support NFRHT enhancements comparable to those of bulk samples when the gap size is smaller than 
the film thickness \cite{Song2015b}. This is possible due to the fact that in these materials the
NFRHT is dominated by the contribution of surface phonon polaritons (SPhPs) whose penetration depth
is comparable to the gap size. In the case of doped semiconductors, like InSb, the NFRHT is also
dominated by surface electromagnetic modes due to either SPhPs or surface plasmon polaritons (SPPs).
Therefore, similar NFRHT enhancements are also expected in thin films made of these materials. Moreover,
since the application of a magnetic field in doped semiconductors results in the generation of 
hyperbolic modes, which have propagating character even in the near-field regime and therefore propagate
further inside the materials, one could expect more drastic magnetic-field effects in thin-film 
structures of doped semiconductors than in their bulk counterparts. Motivated by this idea, we 
study in this section the magnetic-field dependence of the radiative heat transfer between structures 
containing thin films of MO materials. In particular, we investigate the system of 
Fig.~\ref{fig-system}(b), which consists of an InSb plate and an InSb thin film of thickness $t$ 
deposited on a semi-infinite Au layer. For the sake of concreteness, we shall assume that the 
external field is applied perpendicularly to the plates, see Fig.~\ref{fig-system}(b). Thus,
as in the previous section, the magnetic-field-dependent permittivity tensor of InSb is given
by Eq.~(\ref{perm-tensor-Hz}).

\begin{figure}[t]
\includegraphics[width=0.85\columnwidth,clip]{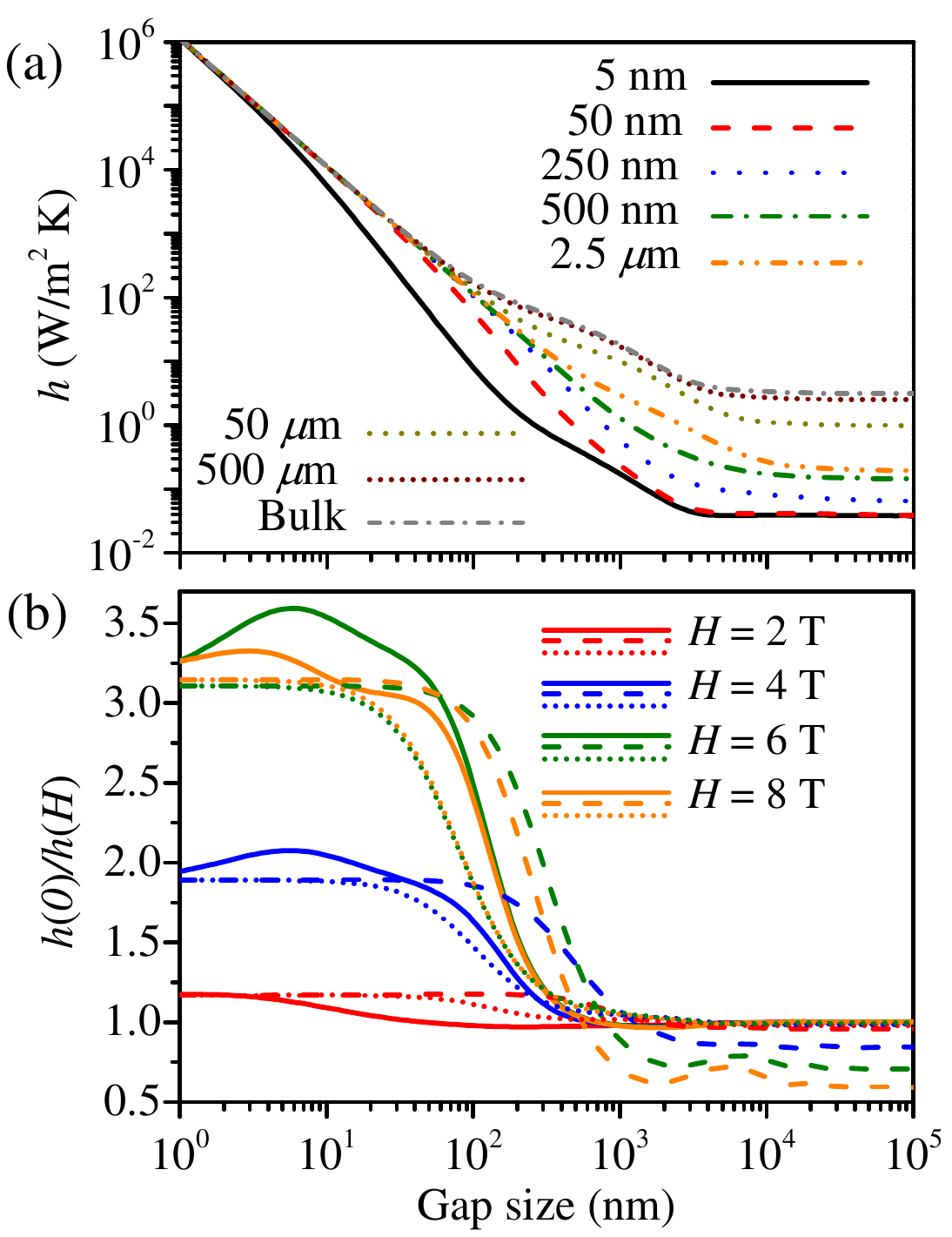}
\caption{(a) Heat transfer coefficient as a function of the gap size for the system of 
Fig.~\ref{fig-system}(b) in the absence of an external magnetic field. The different curves
correspond to different values of the InSb layer thickness, $t$. (b) Ratio of the zero-field heat 
transfer coefficient and the heat transfer coefficient at a given magnetic field for the system of 
Fig.~\ref{fig-system}(b) as a function of the gap size. The different curves correspond to different 
values of the magnitude of the field and to different values of the InSb film thickness: 5 nm (solid 
lines), $2.5$ $\mu$m (dashed lines), and $500$ $\mu$m (dotted lines).}
\label{fig-thin-film1}
\end{figure}

Let us start by analyzing the system of Fig.~\ref{fig-system}(b) in the absence of an external
magnetic field. In Fig.~\ref{fig-thin-film1}(a) we present the results for the heat transfer 
coefficient as a function of the gap size for a variable thickness of the InSb layer deposited 
on Au ranging from 5 nm to a bulk sample. As one can see, in the near-field regime (mainly below 
1 $\mu$m), when the gap size is smaller than the InSb layer thickness, the value of the heat transfer 
coefficient is the same, irrespective of layer thickness. This is exactly the same behavior that 
takes place in the case of polar dielectrics (like SiO$_2$, SiN, etc.) 
\cite{Song2015b,Biehs2007,Basu2009,Francoeur2011} and it has the same physical origin. This behavior 
is due to the fact that the NFRHT is dominated by cavity surface modes, in this case SPPs and SPhPs 
\cite{Moncada-Villa2015}, whose penetration depth scales with the gap size and, therefore, they 
become more are more confined as the gap is reduced. This explains the thickness independence of 
the NFRHT when the gap is sufficiently small (smaller than the layer thickness).

We turn now to discuss how the presence of an external magnetic field alters the NFRHT in these
multilayer structures. In Fig~\ref{fig-thin-film1}(b) we present the results for the ratio
$h(0)/h(H)$ of the heat transfer coefficient without and with a field applied as a function
of the gap size for different values of a perpendicular magnetic field. In particular, we 
present results for three different values of the InSb layer thickness: $t=5$ nm, a very thin
film, $t=2.5$ $\mu$m, an intermediate thickness, and $t=500$ $\mu$m, which basically corresponds
to a bulk layer. The first thing to notice is that in most cases, irrespective of the gap size,
InSb layer thickness, and magnitude of the field, the radiative heat conductance tends to be 
decreased by the field, as compared to the zero-field case. This is, in particular, what occurs
both in the very thin and in the bulk case (this latter situation was extensively analyzed in
Ref.~\cite{Moncada-Villa2015}) both in the near- and in the far-field regime. Let us recall 
that this field-induced reduction is mainly due to the appearance of hyperpolic modes in certain 
frequency regions, modes that are less effective transferring the heat across the gap than 
the surface modes that dominate the NFRHT in the absence of field. Concerning the thickness 
dependence of the heat transfer coefficient, there are two salient effects. First, the largest
field-induced reductions of the NFRHT occur for the thinnest case. This is expected since 
the hyperbolic modes that appear at finite field have long penetration depths (much longer than
surface modes) and, therefore, are more sensitive to the presence of the Au substrate. In practice,
this means that the substrate reduces the probability of the hyperbolic modes and, in turn, their 
contribution to the heat transfer. The second interesting effect is the fact that in the case of 
intermediate thickness values, see curves for $t=2.5$ $\mu$m, there is a gap range (between 
1 and 10 $\mu$m) where the NFRHT is actually enhanced by the field (i.e., the ratio $h(0)/h(H)$ 
is smaller than 1), and this enhancement persists in the far-field regime. This behavior is at 
variance with the bulk case and we shall try to explain its physical origin in what follows,
focusing on the near-field regime.

\begin{figure}[t]
\includegraphics[width=\columnwidth,clip]{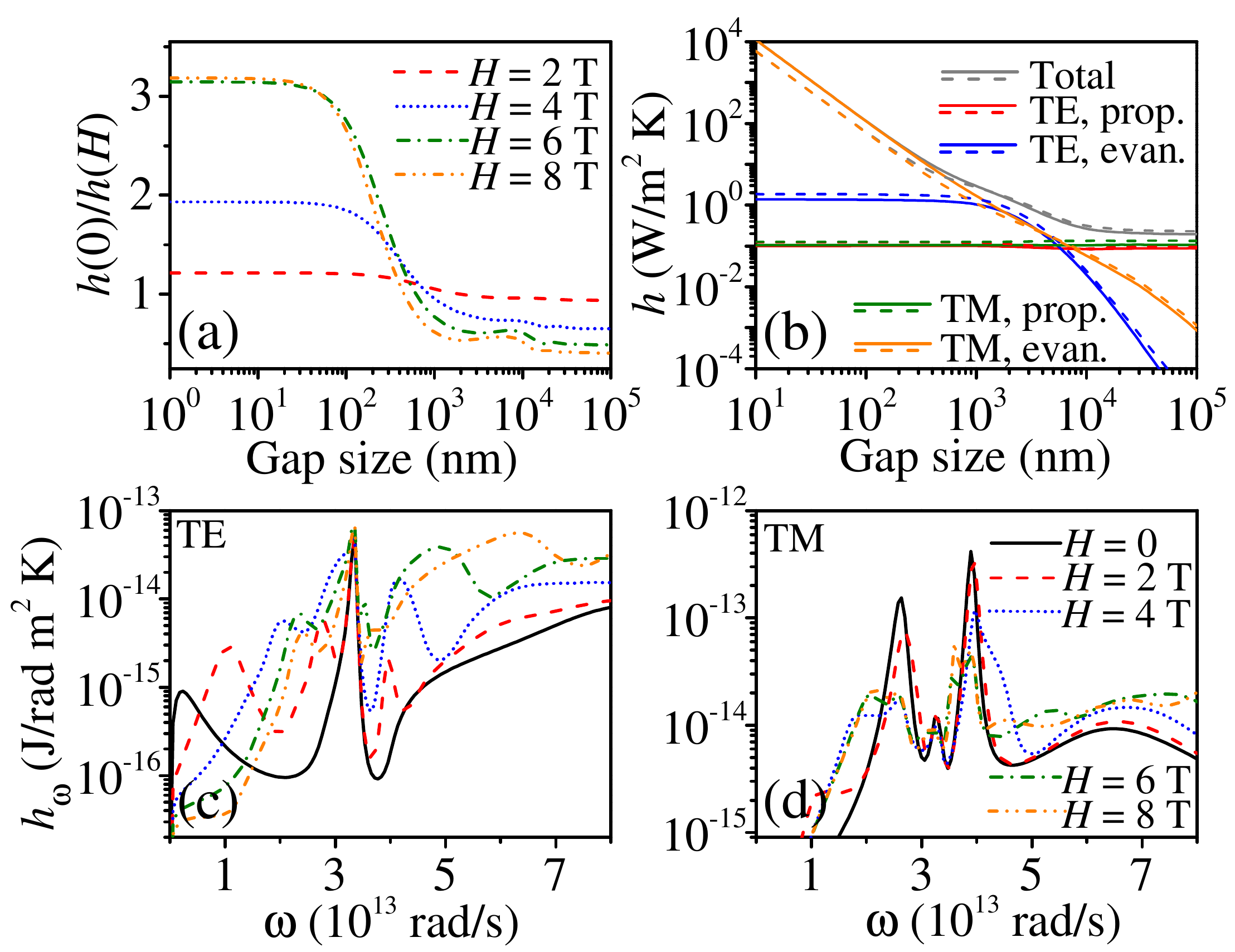}
\caption{(a) Ratio of the zero-field heat transfer coefficient and the heat transfer
coefficient for the system of Fig.~\ref{fig-system}(b) as a function of the gap size and for
InSb layer thickness $t=2.5$ $\mu$m. The different curves correspond to different values of the 
magnitude of the field and the results were computed with the diagonal approximation: 
$r^{s,p}_{ij} = r^{p,s}_{ij} = 0$ in Eq.~\eqref{refl-mat}. Notice that these results accurately
reproduce the exact results shown in Fig~\ref{fig-thin-film1}(b). (b) The corresponding heat transfer
coefficient computed with the diagonal approximation as a function of the gap size for two values of 
the magnetic field: $H=0$ T (solid lines) and $H=4$ T (dashed lines). We show the total contribution
to the heat transfer coefficient as well as the different individual contributions from TE and TM
modes, both propagating and evanescent. Panels (c) and (d) show the spectral heat flux calculated
with the diagonal approximation as a function of frequency for TE ($s$-polarized) and TM ($p$-polarized) 
waves, respectively, for a gap $d = 1$ $\mu$m and different values of the magnitude of the external 
(perpendicular) field.}
\label{fig-thin-film2}
\end{figure}

The first step to understand this field-induced enhancement is to examine the validity of the diagonal
approximation discussed in the previous section where the polarization conversion is ignored
by setting $r^{s,p}_{ij} = r^{p,s}_{ij} = 0$ in Eq.~\eqref{refl-mat}. As we show in 
Fig.~\ref{fig-thin-film2}(a), this approximation accurately reproduces the exact results shown
in Fig~\ref{fig-thin-film1}(b) for the field dependence of the heat transfer coefficient for a 
thickness $t=2.5$ $\mu$m. Thus, we shall use this approximation to shed light on the origin of the
field-induced enhancement. Within this diagonal approximation we can unambiguously define the contribution 
of TE and TM modes, both propagating and evanescent. We show all those contributions in Fig.~\ref{fig-thin-film2}(b)
as a function of the gap size for $t=2.5$ $\mu$m and two values of the field: 0 and 4 T. 
From these results we can learn two basic things. First, the field-induced enhancement of the NFRHT 
occurs in a gap region in which the contributions of evanescent TM and TE modes are comparable, i.e., 
for separations were surface modes do not exclusively dominate the heat transfer. Second, the enhancement 
of the NFRHT due to the external field in this region is due to an increase in the contribution of 
evanescent TE modes. Very much like in the case studied in the previous section, the enhanced 
probability of evanescent TE modes upon the application of the field is due to a reduction of the
impedance mismatch between InSb and the vacuum gap in frequency regions where the dielectric component
$\epsilon_{xx}$ exhibits a vanishing real part. For completeness, we also show
in Fig~\ref{fig-thin-film2}(c,d) the corresponding spectral heat flux for various field values 
for the case $t=2.5$ $\mu$m and $d=1$ $\mu$m. As one can see, the TE spectrum is much more
sensitive to the application of an external field than the corresponding TM modes.

\section{Anisotropic thermal magnetoresistance} \label{sec-ATMR}

In Ref.~\cite{Abraham-Ekeroth2018} it was predicted that a huge anisotropic thermal magnetoresistance 
(ATMR) in the near-field radiative heat transfer between MO particles when the direction 
of an external magnetic field is changed with respect to the heat transport direction. This phenomenon,
which is the thermal analog of the famous anisotropic magnetoresistance (AMR) in spintronics 
\cite{Zutic2004}, was illustrated in the case of two submicron InSb spherical particles. The ATMR was 
first hinted in our work \cite{Moncada-Villa2015} where we studied the magnetic field dependence of
the radiative heat transfer between two identical parallel plates made of doped semiconductors (InSb
and Si). There, we noticed the significant difference in the NFRHT between applying the magnetic field 
perpendicular or parallel to the plates, but we did not analyze systematically the impact of the field
direction on the radiative heat conductance. We shall fill this gap in this section by analyzing the 
NFRHT in the system of Fig.~\ref{fig-system}(c) where two identical InSb infinite plates are subjected 
to a magnetic field forming an angle $\theta_H$ with the transport direction. In particular, we shall 
investigate the change in the heat transfer coefficient for a given value of the magnitude of the external 
field upon changing its orientation relative to the transport direction. 

\begin{figure}[t]
\includegraphics[width=\columnwidth,clip]{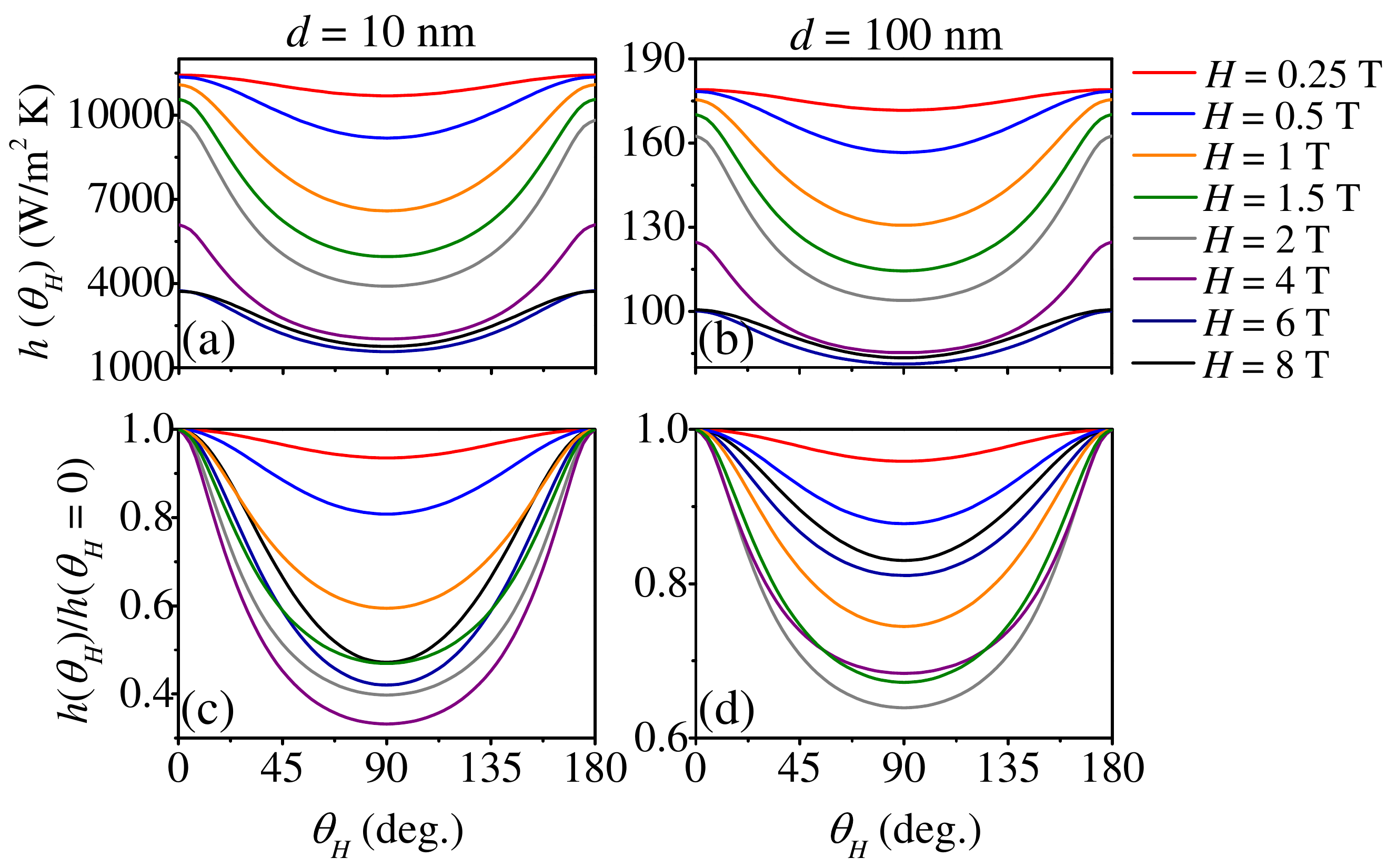}
\caption{(a) Heat transfer coefficient for the system of Fig.~\ref{fig-system}(c) 
as a function of the angle $\theta_H$ between the external field and the transport direction 
for different values of the magnitude of the magnetic field for a gap size $d=10$ nm. (b) The 
same as in panel (b), but for $d=100$ nm. (c,d) The same as in panels (a) and (b), respectively, 
but now the heat transfer coefficient is normalized by its value for $\theta_H$ = 0.}
\label{fig-ATMR}
\end{figure}

In Fig.~\ref{fig-ATMR} we present the results for the dependence of the heat transfer coefficient on the
angle $\theta_H$ between the external magnetic field and the transport direction, see Fig.~\ref{fig-system}(c), 
for different values of the field magnitude and two different values of the gap size in the near-field
regime, 10 and 100 nm. In this figure we show the results for both the absolute value of the heat 
transfer coefficient, panel (a) and (b), and the ratio between its value at given angle $\theta_H$ and 
its value at $\theta_H = 0$ when the field is perpendicular to the plates, panels (c) and (d). As one can see, 
the heat transfer coefficient is strongly modulated by the field direction and it is symmetric around 
$\theta_H = 90^{\rm o}$. In all cases, the conductance is maximum at $\theta_H = 0^{\rm o}$ (when the field
is perpendicular to the plates) and it reaches a minimum at $\theta_H = 90^{\rm o}$ (when the field
is parallel to the plates). More importantly, the ATMR ratio, defined as $h(\theta_H)/h(\theta_H=0)$,
reaches a minimum, for example, for 1 T of $\sim 0.6$ for $d=10$ nm and of $\sim 0.75$ for $d=100$ nm.
In terms of a thermal resistance per unit of area, $R = 1/h$, these ATMR ratios imply relative changes 
$[R(\theta_H) - R(\theta_H = 0)]/R(\theta_H = 0)$ of approximately 66\% and 33\%, respectively. 
These values are indeed remarkable when one compares them with the 1\% relative change in the resistance 
of spintronic devices for similar fields \cite{OHandley2000}. It is also worth noticing that the heat 
transfer coefficient is significantly modulated for moderate fields of 0.25 T (see red curves in 
Fig.~\ref{fig-ATMR}). Notice also the ATMR ratio does not decrease monotonically with the magnitude of 
field in the high-field range (above 1 T).

\begin{figure}[t]
\includegraphics[width=\columnwidth,clip]{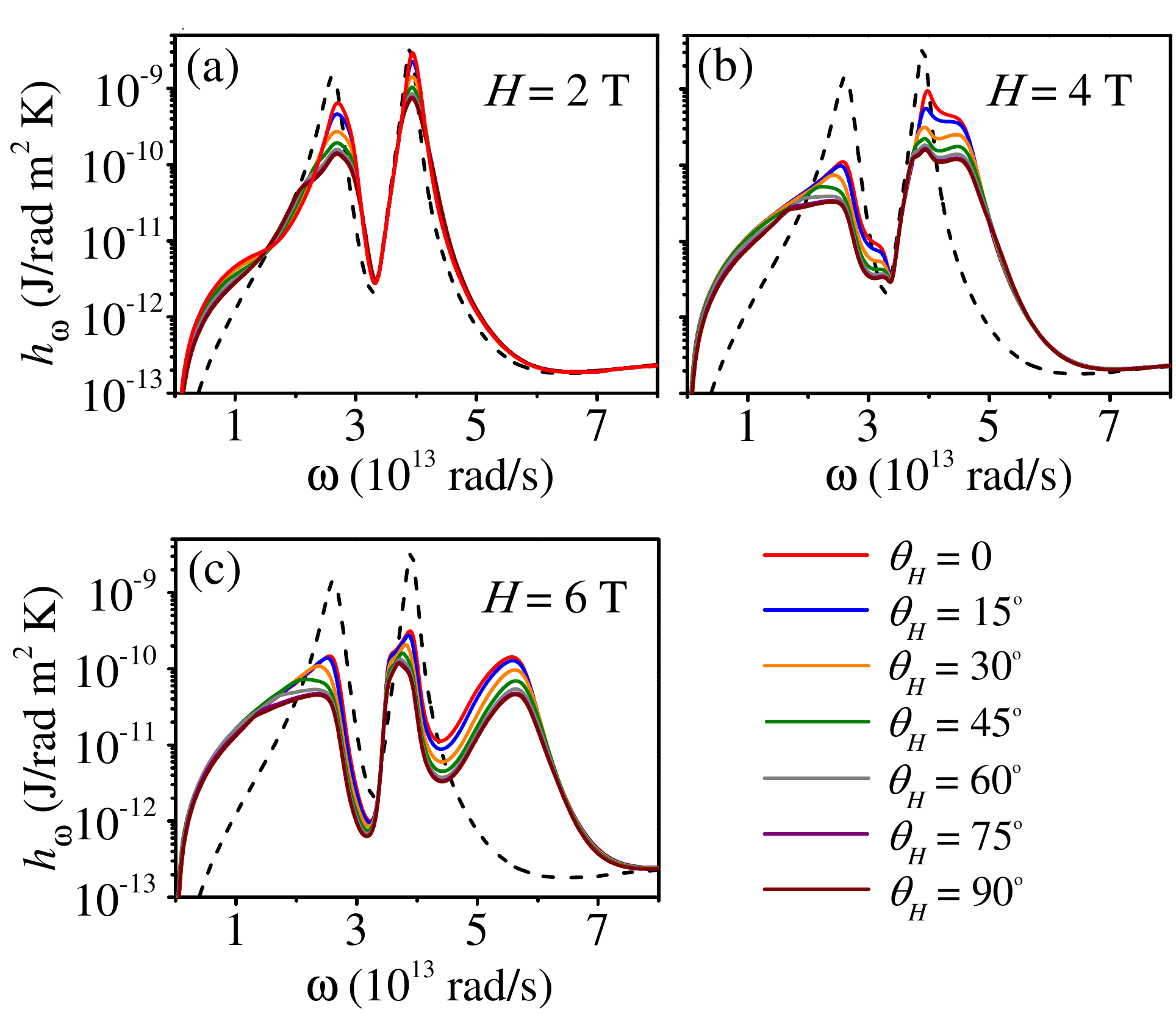}
\caption{Spectral heat flux for the system of Fig.~\ref{fig-system}(c) as a 
function of the frequency and the angle $\theta_H$ between the external field and the transport 
direction, for a gap of $d=10$ nm. Panels (a), (b) and (c) correspond to a magnetic field magnitude 
of 2 T, 4 T, and 6 T, respectively. The dashed lines correspond to the zero-field case.}
\label{fig-ATMR-spectral}
\end{figure}

The physical explanation of the angular dependence of the heat transfer coefficient in the near-field
regime is quite complex. Some insight can be gained by analyzing the spectral heat flux. We
show the results for this quantity in Fig.~\ref{fig-ATMR-spectral} for a gap of $d=10$ nm and three
different values of the magnitude of the external field (2, 4, and 6 T). The different curves 
correspond to different value of the angle $\theta_H$ and for comparison, we also present the 
results in the absence of an external field (dashed lines). Notice that the by increasing $\theta_H$ 
from 0$^{\rm o}$ to 90$^{\rm o}$, the main effect is reduction of the height of the main peaks in 
the spectra, while the position of those peaks, which are due to the different surface modes present 
in this system, remains more or less unchanged. A simple interpretation or explanation of these
spectra is very difficult due to the fact that for $\theta_H \neq 0$ the radiative heat transfer
becomes anisotropic in $k$-space, see Eq.~(\ref{eq-net-Q}), meaning that it does not longer depend 
only on the modulus of the parallel wave vector, $k$, but rather on the exact direction ${\bf k} = (k_x,k_y)$.
As we showed in Ref.~\cite{Moncada-Villa2015}, when the field is parallel to the plates, there 
are a variety of electromagnetic modes that contribute to the NFRHT depending of ${\bf k}$ direction.
In that particular configuration, surface modes originating from both plasmon and phonon polaritons 
compete with hyperbolic modes. In particular, the dispersion relation of the surface modes is very 
sensitive to the external magnetic field, which leads to an even more drastic reduction of the NFRHT 
than in the case of a field perpendicular to the plates and explains why the heat transfer coefficient
reaches a minimum at $\theta_H = 90^{\rm o}$. 

\section{Conclusions} \label{sec-conclusions}

As explained in the introduction, the application of a static magnetic field in combination with
MO materials has emerged as a promising way to actively tune and control NFRHT. However, this
generic idea has still to be demonstrated experimentally and most of the related thermomagnetic effects
predicted in the last few years deal with systems that are extremely challenging for the experiment.
For this reason, we have focused on this work on putting forward a number of novel physical effects
related to the NFRHT between planar layered systems containing MO materials that are amenable to measurement
with the existent experimental techniques. In particular, we have discussed three basic phenomena.
First, we have shown that, contrary to what have been reported so far, it is possible to use a 
magnetic field to increase the NFRHT. We have demonstrated this possibility with the analysis of 
an asymmetric system consisting of two parallel plates made of InSb and Au. We have attributed the
field-induced enhancement of the NFRHT to the corresponding increase in the contribution of evanescent 
TE waves due to the reduction of the impedance mismatch between InSb and the vacuum gap in frequency
regions where the real part of the InSb refractive index tends to vanish.

The second physical problem that we have explored in this work is the magnetic field dependence of
the NFRHT in layered structures of MO materials containing thin films. To be precise, we have 
studied the radiative heat transfer between an InSb infinite plate and a bilayer structure comprising
an InSb thin film deposited on a Au substrate. We have shown that depending on the thickness of the InSb
film one can further reduce the NFRHT upon the application of an external field perpendicular to the 
plate surfaces. We have also found that in the case of films with intermediate thicknesses it is
possible to enhance the NFRHT in a certain gap range, contrary to what happens in bulk counterparts. 
We have shown that this peculiar effect also originates from the field-induced enhancement of the 
probability of evanescent TE modes. 

Finally, we have analyzed systematically the phenomenon of anisotropic thermal magnetoresistance (ATMR)
in the case of two parallel plates made of InSb, extending so our partial analysis of Ref.~\cite{Moncada-Villa2015}.
We have shown that the NFRHT is very sensitive to the orientation of the external field, with respect 
to the transport direction, and its modulation only requires moderate magnetic fields. Moreover, the 
amplitudes of the field modulation of the radiative heat conductance in this simple situation turn out 
to be orders of magnitude larger than in the case of the electronic analog in the context of spintronic 
devices.

So, in short, the different phenomena predicted in this work illustrate once more the rich thermal 
radiation physics that one can encounter when combining an external magnetic field and MO materials.
Moreover, they show that the application of an external magnetic field enables to actively control 
NFRHT in a variety of ways and we are convinced that this work will motivate the realization of 
experiments to test all these different ideas.

\acknowledgments

We thank Antonio I. Fern\'andez-Dom\'{\i}nguez for fruitful discussions. J.C.C.\ acknowledges funding 
from the Spanish Ministry of Economy and Competitiveness (MINECO) (contract No.\ FIS2017-84057-P).

\end{document}